\documentclass[preprint,showpacs,aps,draft]{revtex4}

\usepackage{graphicx}
\usepackage{dcolumn}
\usepackage{bm}
\usepackage{amsmath}
\makeatletter

\newcommand{\Rmnum}[1]{\expandafter\@slowromancap\romannumeral #1@}
\makeatother

\begin{document}

\title{Decoherence induced spontaneous symmetry breaking}

\author{G. Karpat}

\author{Z. Gedik}

\affiliation{Faculty of Engineering and Natural Sciences, Sabanci University, Tuzla, Istanbul 34956, Turkey}

\date{\today}

\begin{abstract}
We study time dependence of exchange symmetry properties of Bell states when two qubits
interact with local baths having identical parameters. In case of classical noise, we consider
a decoherence Hamiltonian which is invariant under swapping the first and second qubits.
We find that as the system evolves in time, two of the three symmetric Bell states preserve
their qubit exchange symmetry with unit probability, whereas the symmetry of the remaining
state survives with a maximum probability of $ 0.5 $ at the asymptotic limit. Next, we examine the exchange symmetry
properties of the same states under local, quantum mechanical noise which is modeled by two
identical spin baths. Results turn out to be very similar to the classical case. We identify
decoherence as the main mechanism leading to breaking of qubit exchange symmetry.
\end{abstract}


\maketitle

\section{Introduction}

Since the early days of quantum mechanics, it has been known that certain quantum states have a mysterious
non-local behavior [1]. The phenomenon responsible for these non-local correlations among the subsystems
of a composite quantum system is called entanglement [2]. Quantum entanglement, having no
classical counterpart, is believed to be one of the characteristic features of quantum mechanics.
Besides its foundational importance for the quantum theory, entanglement is also considered as the
resource of quantum computation, quantum cryptography and quantum information processing [3].
In recent years, it has been extensively studied with various motivations [4]. However, entanglement of
quantum systems, as all other quantum traits, is very fragile when they are exposed to external disturbances,
which is inevitably the case in real world situations.

Decoherence, the process through which quantum states lose their phase relations irreversibly due
to interactions with the environment, is crucial for understanding the emergence of classical behavior in quantum
systems [5]. It also presents a major challenge for the realization of quantum information processing protocols
since protection of non-local correlations against undesirable external disturbances is essential for the reliability
of such protocols. Consequently, understanding the decoherence effect of the environment on entangled systems is
an important issue. This problem has been currently addressed in literature, considering both local and collective
interactions of qubits and qutrits with the environment. While some authors examined the effects of classical stochastic
noise fields [6-9], others studied the same problem for large spin environments [10-15].

In this work, we focus on a different aspect of a decoherence process of entangled states. Certain two-qubit entangled states
have the property that they remain unchanged under the exchange of two qubits. We will concentrate on a decoherence model which
also has an exchange symmetry, i.e., having a Hamiltonian invariant upon swapping the first and second qubits. Our goal is
to understand how the exchange symmetry properties of symmetric pure states alter as the quantum system evolves in time
for a symmetric Hamiltonian which embodies the effect of local and identical noise fields on qubits. More specifically, we
will investigate the exchange symmetry properties of three of the four Bell states. Bell states are defined as maximally
entangled quantum states of two-qubit systems and given as
\begin{eqnarray}
\vert B_{1} \rangle = \frac{1}{\sqrt{2}}(\vert 00\rangle +\vert 11\rangle),
\label{eq:1}
\end{eqnarray}
\begin{eqnarray}
\vert B_{2} \rangle = \frac{1}{\sqrt{2}}(\vert 00\rangle -\vert 11\rangle),
\label{eq:2}
\end{eqnarray}
\begin{eqnarray}
\vert B_{3} \rangle = \frac{1}{\sqrt{2}}(\vert 01\rangle +\vert 10\rangle),
\label{eq:3}
\end{eqnarray}
\begin{eqnarray}
\vert B_{4} \rangle = \frac{1}{\sqrt{2}}(\vert 01\rangle -\vert 10\rangle).
\label{eq:4}
\end{eqnarray}
We will only consider the first three of these states which are symmetric under exchange operation.
However, our discussion can be extended to include anti-symmetric states like $ \vert B_{4} \rangle $.
The first three Bell states are among the symmetric pure two-qubit states which can be represented in the
most general case by the density matrix
\begin{eqnarray}
\rho_{sym} =
\left(\begin{array}{cccc}
\vert a \vert^{2} & a c^{*} & a c^{*} & a b^{*} \\
c a^{*} & \vert c \vert^{2} & \vert c \vert^{2} & c b^{*} \\
c a^{*} & \vert c \vert^{2} & \vert c \vert^{2} & c b^{*} \\
b a^{*} & b c^{*} & b c^{*} & \vert b \vert^{2} \\
\end{array}\right),
\label{eq:5}
\end{eqnarray}
where $ \vert a \vert^{2}+ 2\vert c \vert^{2}+\vert b \vert^{2}=1 $.
After classical noise calculations, we will briefly discuss the exchange symmetry properties of the same states
for local and quantum mechanical noise which is modeled via two identical large spin environments.

\section{Local Classical Noise}

We assume that the two qubits are interacting with separate baths locally and
the initial two-qubit system is not entangled with the local baths.
The model Hamiltonian we consider was first introduced and studied by Yu and Eberly [6]
and can be thought as the representative of the class of interactions which generate a
pure dephasing process that is defined as
\begin{eqnarray}
H(t)= -\frac{1}{2} \mu [n_{A}(t)(\sigma_{z}\otimes I)+ n_{B}(t)(I\otimes\sigma_{z})],
\label{eq:6}
\end{eqnarray}
where we take $ \hbar=1 $ and $ \sigma_{z} $ is the Pauli matrix
\begin{eqnarray}
\sigma_{z} =
\left(\begin{array}{cc}
1 & 0 \\
0 & -1 \\
\end{array}\right).
\label{eq:7}
\end{eqnarray}
Here $ \mu $ is the gyromagnetic ratio and $n_{A}(t),n_{B}(t) $ are stochastic noise fields that lead to statistically
independent Markov processes satisfying
\begin{eqnarray}
\langle n_{i}(t) \rangle = 0,
\label{eq:8}
\end{eqnarray}
\begin{eqnarray}
\langle n_{i}(t)n_{i}(t') \rangle = \frac{\Gamma_{i}}{\mu^{2}} \delta(t-t'),
\label{eq:9}
\end{eqnarray}
where $ \langle \cdots \rangle $ stands for ensemble average and $ \Gamma_{i} $(i=A,B) are the damping rates associated
with the stochastic fields $ n_{A}(t) $ and $ n_{B}(t) $.

The time evolution of the system's density matrix can be obtained as
\begin{eqnarray}
\rho(t)= \langle U(t) \rho(0) U^{\dagger}(t) \rangle,
\label{eq:10}
\end{eqnarray}
where ensemble averages are evaluated over the two noise fields $ n_{A}(t) $ and $ n_{B}(t) $
and the time evolution operator, $ U(t) $, is given by
\begin{eqnarray}
U(t)= \exp [-i \int_0^t \! dt' H(t') \ ].
\label{eq:11}
\end{eqnarray}
The resulting density matrix in the product basis $\lbrace\vert 00\rangle,\vert 01\rangle,\vert 10\rangle,\vert 11\rangle \rbrace$ can be written as
\begin{eqnarray}
\rho(t)=
\left(\begin{array}{cccc}
\rho_{11} & \rho_{12} \gamma_{B} & \rho_{13} \gamma_{A} & \rho_{14} \gamma_{A}\gamma_{B} \\
\rho_{21} \gamma_{B} & \rho_{22} & \rho_{23} \gamma_{A}\gamma_{B}& \rho_{24} \gamma_{A} \\
\rho_{31} \gamma_{A} & \rho_{32} \gamma_{A}\gamma_{B} & \rho_{33} & \rho_{34} \gamma_{B} \\
\rho_{41} \gamma_{A}\gamma_{B} & \rho_{42} \gamma_{A} & \rho_{43} \gamma_{B} & \rho_{44} \\
\end{array}\right),
\label{eq:12}
\end{eqnarray}
where $ \rho_{ij} $ stands for the elements of the initial density matrix, $ \rho(0), $ and $ \gamma_{A} $, $\gamma_{B} $ are given by
\begin{eqnarray}
\gamma_{A}(t)= e^{ -t \Gamma_{A} /2}, \qquad \gamma_{B}(t)= e^{ -t \Gamma_{B} /2}.
\label{eq:13}
\end{eqnarray}
For our purposes, we want our two local baths to be identical in a sense that they have the same dephasing rate $ \Gamma $.
Therefore, we let $\Gamma_{A}=\Gamma_{B}=\Gamma $. The resulting density matrix of the system with the consideration of identical baths
is now given by
\begin{eqnarray}
\rho(t)=
\left(\begin{array}{cccc}
\rho_{11} & \rho_{12} \gamma & \rho_{13} \gamma & \rho_{14} \gamma^{2} \\
\rho_{21} \gamma & \rho_{22} & \rho_{23} \gamma^{2} & \rho_{24} \gamma \\
\rho_{31} \gamma & \rho_{32} \gamma^{2} & \rho_{33} & \rho_{34} \gamma \\
\rho_{41} \gamma^{2} & \rho_{42} \gamma & \rho_{43} \gamma & \rho_{44} \\
\end{array}\right),
\label{eq:14}
\end{eqnarray}
where $ \gamma_{A}=\gamma_{A}=\gamma$.

\section{Operator-Sum Representation of Decoherence}

To examine the symmetry properties, we need to express the dynamical evolution of
$ \rho(t) $ in terms of quantum operations. The decoherence process
of our quantum system can be regarded as a completely
positive linear map $ \Phi(\rho) $, that takes an initial state $ \rho(0) $
and maps it to some final state $ \rho(t) $ [3].
For every completely positive linear map there exists an operator-sum representation
which is known as Kraus representation [16-18]. The effect of the map is given by
\begin{eqnarray}
\rho(t)= \Phi(\rho(0))= \sum_{\mu=1}^{N} K_{\mu}(t)\rho(0)K_{\mu}^{\dagger}(t),
\label{eq:15}
\end{eqnarray}
where $  K_{\mu} $ are the Kraus operators which satisfy the unit trace condition
\begin{eqnarray}
\sum_{\mu=1}^{N} K_{\mu}^{\dagger}(t)K_{\mu}(t)=I.
\label{eq:16}
\end{eqnarray}

The Kraus operator approach provides an elegant way to study the decoherence process.
In order to describe the internal decoherence dynamics of the system, all we need to
know is the Kraus operator set which inherently contains the entire information about
environment. The operator-sum representation of our completely positive linear map, $ \Phi(\rho) $,
which reflects the effect of the stochastic process, can be obtained by studying the
mapping called Choi-Jamiolkowski isomorphism [17-18]. In our investigation, it turns out
that the effect of the mapping, $ \Phi(\rho) $, on the two-qubit system can be
expressed by a set of four Kraus operators as
\begin{eqnarray}
K_{1}= \frac{1}{\sqrt{2}}
\left(\begin{array}{cccc}
-\omega(t) & 0 & 0 & 0 \\
0 & 0 & 0 & 0 \\
0 & 0 & 0 & 0 \\
0 & 0 & 0 & \omega(t) \\
\end{array}\right),
\label{eq:17}
\end{eqnarray}
\begin{eqnarray}
K_{2}= \frac{1}{\sqrt{2}}
\left(\begin{array}{cccc}
0 & 0 & 0 & 0 \\
0 & -\omega(t) & 0 & 0 \\
0 & 0 & \omega(t) & 0 \\
0 & 0 & 0 & 0 \\
\end{array}\right),
\label{eq:18}
\end{eqnarray}
\begin{eqnarray}
K_{3}= \frac{1}{2}
\left(\begin{array}{cccc}
\alpha(t) & 0 & 0 & 0 \\
0 & -\alpha(t) & 0 & 0 \\
0 & 0 & -\alpha(t) & 0 \\
0 & 0 & 0 & \alpha(t) \\
\end{array}\right),
\label{eq:19}
\end{eqnarray}
\begin{eqnarray}
K_{4}= \frac{1}{2}
\left(\begin{array}{cccc}
\beta(t) & 0 & 0 & 0 \\
0 & \beta(t) & 0 & 0 \\
0 & 0 & \beta(t) & 0 \\
0 & 0 & 0 & \beta(t) \\
\end{array}\right),
\label{eq:20}
\end{eqnarray}
where $ \omega(t) $, $ \alpha(t) $ and $ \beta(t) $ are given by
\begin{eqnarray}
\omega(t)=\sqrt{1-\gamma(t)^{2}}, \qquad \alpha(t)=\gamma(t)-1, \qquad \beta(t)=\gamma(t)+1.
\label{eq:21}
\end{eqnarray}
Since different environmental interactions may result in the same dynamics on the system,
the operator-sum representation of a quantum process is not unique. The collective
action of our set of the four Kraus operators $\lbrace K_{1},K_{2},K_{3},K_{4} \rbrace$
on the density matrix of the two-qubit quantum system are equivalent to the collective
action of another set of Kraus operators $\lbrace E_{1},E_{2},E_{3},E_{4} \rbrace$
if and only if there exists complex numbers $ u_{ij} $ such that $ E_{i}=\sum_{j}u_{ij}K_{j} $
where $ u_{ij} $ are the elements of a $4\times 4$ unitary matrix [3]. The unitary freedom will
provide us an easy way to introduce exchange symmetry condition.

\section{Exchange Symmetry of Bell States Under Decoherence}

In order to analyze the exchange symmetries of symmetric Bell states
$ \vert B_{1} \rangle,\vert B_{2} \rangle$ and  $\vert B_{3} \rangle
$, we exploit the unitary freedom on the operator-sum
representation. Consider the most general $4\times 4$ unitary matrix
with complex elements
\begin{eqnarray}
U =
\left(\begin{array}{cccc}
u_{11} & u_{12} & u_{13} & u_{14} \\
u_{21} & u_{22} & u_{23} & u_{24} \\
u_{31} & u_{32} & u_{33} & u_{34} \\
u_{41} & u_{42} & u_{43} & u_{44} \\
\end{array}\right),
\label{eq:22}
\end{eqnarray}
where $ U^{\dagger}U=I $.

The mapping described by the Kraus operators $\lbrace
K_{1},K_{2},K_{3}, K_{4} \rbrace $ , via unitary freedom, is
equivalent to the mappings described by the following four Kraus
operators
\begin{align}
\begin{split}
E_{\mu} =
\mathrm{Diag}[-\frac{\omega u_{\mu 1}}{\sqrt{2}}+\frac{\alpha u_{\mu 3}}{2}+\frac{\beta u_{\mu 4}}{2},-\frac{\omega u_{\mu 2}}{\sqrt{2}}-\frac{\alpha u_{\mu 3}}{2}+\frac{\beta u_{\mu 4}}{2}, \\
\frac{\omega u_{\mu 2}}{\sqrt{2}}-\frac{\alpha u_{\mu 3}}{2}+\frac{\beta u_{\mu 4}}{2},\frac{\omega u_{\mu 1}}{\sqrt{2}}+\frac{\alpha u_{\mu 3}}{2}+\frac{\beta u_{\mu 4}}{2}],
\label{eq:23}
\end{split}
\end{align}
where $ \mu =1,2,3,4. $

\subsection{Exchange Symmetries of  $ \vert B_{1} \rangle $ and $\vert B_{2} \rangle $}

Having calculated all possible Kraus operator sets, we are in a position to evaluate the
possible final states when the initial state is $ \vert B_{1} \rangle $ or $\vert B_{2} \rangle $.
The density matrices of possible final states are obtained as
\begin{eqnarray}
\rho^{B_{1}}_{\mu}(t)=  \frac{E_{\mu}(t)\rho^{B_{1}}(0)E_{\mu}^{\dagger}(t)}{Tr(E_{\mu}(t)\rho^{B_{1}}(0)E_{\mu}^{\dagger}(t))}, \qquad \rho^{B_{2}}_{\mu}(t)=  \frac{E_{\mu}(t)\rho^{B_{2}}(0)E_{\mu}^{\dagger}(t)}{Tr(E_{\mu}(t)\rho^{B_{2}}(0)E_{\mu}^{\dagger}(t))},
\label{eq:24}
\end{eqnarray}
where $\mu =1,2,3,4 $ and $ \rho^{B_{i}}(0) = \vert B_{i} \rangle \langle B_{i} \vert $ for  $i =1,2$.
The explicit forms of the density matrices $\rho^{B_{1}}_{\mu}(t)$ and $\rho^{B_{2}}_{\mu}(t)$ are given by
\begin{eqnarray}
\rho_{\mu}^{B_{1}}(t)= \frac{1}{\vert e \vert^{2}+\vert f \vert^{2}}
\left(\begin{array}{cccc}
\vert e \vert^{2} & 0 & 0 & ef^{*} \\
0 & 0 & 0 & 0 \\
0 & 0 & 0 & 0 \\
e^{*}f & 0 & 0 & \vert f \vert^{2}  \\
\end{array}\right),
\label{eq:25}
\end{eqnarray}
\begin{eqnarray}
\rho_{\mu}^{B_{2}}(t)= \frac{1}{\vert e \vert^{2}+\vert f \vert^{2}}
\left(\begin{array}{cccc}
\vert e \vert^{2} & 0 & 0 & -ef^{*} \\
0 & 0 & 0 & 0 \\
0 & 0 & 0 & 0 \\
-e^{*}f & 0 & 0 & \vert f \vert^{2}  \\
\end{array}\right),
\label{eq:26}
\end{eqnarray}
where
\begin{eqnarray}
e = (\frac{-\omega u_{\mu 1}}{\sqrt{2}}+\frac{\alpha u_{\mu 3}}{2}+\frac{\beta u_{\mu 4}}{2}), \qquad
f = (\frac{\omega u_{\mu 1}}{\sqrt{2}}+\frac{\alpha u_{\mu 3}}{2}+\frac{\beta u_{\mu 4}}{2}).
\label{eq:27}
\end{eqnarray}

Obviously, the symmetry condition given in Eq. (5) brings no restriction on these density matrices. Thus, it is
guaranteed that the Bell states $ \vert B_{1} \rangle $ and $\vert B_{2} \rangle $ always preserve their
exchange symmetry as they evolve in time under our model Hamiltonian.

\subsection{Exchange Symmetry of  $ \vert B_{3} \rangle $}

The density matrices of the possible final states for $ \vert B_{3} \rangle $ are written as
\begin{eqnarray}
\rho^{B_{3}}_{\mu}(t)=  \frac{E_{\mu}(t)\rho^{B_{3}}(0)E_{\mu}^{\dagger}(t)}{Tr(E_{\mu}(t)\rho^{B_{3}}(0)E_{\mu}^{\dagger}(t))},
\label{eq:28}
\end{eqnarray}
where $\mu =1,2,3,4 $ and $ \rho^{B_{3}}(0) = \vert B_{3} \rangle \langle B_{3} \vert $.
The explicit form of the density matrix $\rho^{B_{3}}_{\mu}(t)$ is
\begin{eqnarray}
\rho^{B_{3}}_{\mu}(t)= \frac{1}{\vert r \vert^{2}+\vert s \vert^{2}}
\left(\begin{array}{lccl}
0 & 0 & 0 & 0 \\
0 & \vert r \vert^{2} & rs^{*} & 0 \\
0 & r^{*}s & \vert s \vert^{2} & 0 \\
0 & 0 & 0 & 0 \\
\end{array}\right),
\label{eq:29}
\end{eqnarray}
where
\begin{eqnarray}
r = (\frac{-\omega u_{\mu 2}}{\sqrt{2}}-\frac{\alpha u_{\mu 3}}{2}+\frac{\beta u_{\mu 4}}{2}), \qquad
s = (\frac{\omega u_{\mu 2}}{\sqrt{2}}-\frac{\alpha u_{\mu 3}}{2}+\frac{\beta u_{\mu 4}}{2}).
\label{eq:30}
\end{eqnarray}

As can be seen from the form of the density matrix of the most general two-qubit symmetric pure state in Eq. (5),
for possible final states to be symmetric we need all non-zero elements of the matrix in Eq. (29) to be equal
to each other, that is, $ r=s $ . This condition can only be satisfied in case of $ u_{\mu2}=0  $. We can immediately conclude that
it is impossible for all of the possible final states to be symmetric since any $4\times 4  $ unitary matrix has to
satisfy the condition that $ \vert u_{12} \vert^{2} + \vert u_{22} \vert^{2} + \vert u_{32} \vert^{2} +
\vert u_{42} \vert^{2} =1 $. Thus, $ \vert B_{3} \rangle $ cannot evolve in time under our model Hamiltonian
in a way that preserves its qubit exchange symmetry with unit probability. In other words, the exchange symmetry of
this two-qubit state has to be broken with some non-zero probability. Considering the symmetry of the initial state
and the Hamiltonian this is a very interesting result. A natural question is the maximum probability of finding a symmetric
possible final state as the system evolves in time. In order to answer this question, we need to
consider three different cases, namely, the cases of having one, two or three symmetric possible final states.

If we assume only one of the possible final states to be symmetric, say the outcome of $E_{1}$ ($ u_{12}=0 $), then the probability of getting a
symmetric output state is given by
\begin{eqnarray}
P_{sym}(t \rightarrow \infty) = \frac{1}{4} \vert u_{13} + u_{14}\vert^{2}.
\label{eq:31}
\end{eqnarray}
If we assume two of the possible final states to be symmetric, say the outcomes of $E_{1}$ and $E_{2}$ ($ u_{12}=0 $, $ u_{22}=0 $), then the probability of having a
symmetric output state is given by
\begin{eqnarray}
P_{sym}(t \rightarrow \infty)= \frac{1}{4} \vert u_{13} + u_{14}\vert^{2} + \frac{1}{4} \vert u_{23} + u_{24}\vert^{2}.
\label{eq:32}
\end{eqnarray}
Finally, if three of the possible final states are symmetric, say the outcomes of $E_{1}$, $E_{2}$ and $E_{3}$ ($ u_{12}=0 $, $ u_{22}=0 $, $ u_{32}=0 $), then the probability
of having a symmetric output state is given by
\begin{eqnarray}
P_{sym}(t \rightarrow \infty)= \frac{1}{4} \vert u_{13} + u_{14}\vert^{2} + \frac{1}{4} \vert u_{23} + u_{24}\vert^{2} + \frac{1}{4} \vert u_{33} + u_{34}\vert^{2}.
\label{eq:33}
\end{eqnarray}
In all of these possible cases, the maximum probability of finding a symmetric final state
turns out to be 0.5.

\section{Local Quantum Noise}
When it comes to modeling the baths as large spin environments, one
of the simplest decoherence models, introduced in [19], is that of
two central spins interacting with N independent spins through the
Hamiltonian [13]
\begin{eqnarray}
H = c_{1z}\sum_{k=1}^{N_{1}} \hbar \omega_{1k} \sigma_{1kz} + c_{2z}\sum_{k=1}^{N_{2}} \hbar \omega_{2k} \sigma_{2kz}.
\label{eq:34}
\end{eqnarray}
This model describes two central spins, with \textit{z}-component operators  $ c_{1z} $ and $ c_{2z} $, coupled to bath spins
represented by $ \sigma_{nkz} $, where $ n=1,2 $ labels the baths and $ k=1,2,3,...,N_{n} $ labels the individual spins.
All spins are assumed to be 1/2 and  $ c_{1z},c_{2z} $ and $\sigma_{nkz} $ denote the corresponding Pauli matrices.
If we assume that the central spins are not entangled with the spin baths at $ t=0 $, the initial state will
be in product form $ \vert \Psi(0)\rangle =  \vert \Psi_{c}(0)\rangle \vert \Psi_{\sigma 1}(0)\rangle \vert \Psi_{\sigma 2}(0)\rangle $ where
\begin{eqnarray}
\vert \Psi_{c}(0) \rangle = ( a_{\uparrow\uparrow} \vert \uparrow\uparrow \rangle + a_{\uparrow\downarrow} \vert \uparrow\downarrow \rangle
+ a_{\downarrow\uparrow} \vert \downarrow\uparrow \rangle + a_{\downarrow\downarrow} \vert \downarrow\downarrow \rangle ),
\label{eq:35}
\end{eqnarray}
with
\begin{eqnarray}
\vert \Psi_{\sigma n}(0) \rangle = \bigotimes_{k=1}^{N_{n}} (\alpha_{nk} \vert \uparrow_{nk}  \rangle + \beta_{nk} \vert \downarrow_{nk}  \rangle ),
\label{eq:36}
\end{eqnarray}
where $ \vert \uparrow_{nk}  \rangle $ and $ \vert \downarrow_{nk}  \rangle $ are eigenstates of $ \sigma_{nkz} $ with eigenvalues +1 and -1,
respectively, and $ \vert \alpha_{nk} \vert^{2} +\vert \beta_{nk} \vert^{2} =1 $.

The reduced density matrix of two central spins at later times will be given by tracing out the bath degrees of freedom from the total density matrix
of the system, $ \rho(t) $, as $ \rho_{c}(t)=Tr_{\sigma}\rho(t) $ where subscript $ \sigma $ means that trace is evaluated by summing over all possible
$ nk $ states and $ \rho(t)= \vert \Psi(t) \rangle  \langle \Psi(t) \vert $. The resulting reduced density matrix in product basis
$ \lbrace \vert\uparrow\uparrow\rangle, \vert\uparrow\downarrow\rangle, \vert\downarrow\uparrow\rangle,\vert\downarrow\downarrow\rangle \rbrace $ is found to be
\begin{eqnarray}
\rho_{c} =
\left(\begin{array}{cccc}
\vert a_{\uparrow\uparrow}\vert^{2} & a_{\uparrow\uparrow} a_{\uparrow\downarrow}^{*} r_{2} & a_{\uparrow\uparrow} a_{\downarrow\uparrow}^{*} r_{1} & a_{\uparrow\uparrow} a_{\downarrow\downarrow}^{*} r_{1} r_{2} \\
a_{\uparrow\uparrow}^{*} a_{\uparrow\downarrow} r_{2}^{*} & \vert a_{\uparrow\downarrow} \vert^{2} & a_{\uparrow\downarrow}a_{\downarrow\uparrow}^{*} r_{1} r_{2}^{*} & a_{\uparrow\downarrow}a_{\downarrow\downarrow}^{*} r_{1} \\
a_{\uparrow\uparrow}^{*}a_{\downarrow\uparrow}r_{1}^{*} & a_{\uparrow\downarrow}^{*}a_{\downarrow\uparrow}r_{1}^{*}r_{2} & \vert a_{\downarrow\uparrow}\vert^{2} & a_{\downarrow\uparrow}a_{\downarrow\downarrow}^{*}r_{2} \\
a_{\uparrow\uparrow}^{*} a_{\downarrow\downarrow}r_{1}^{*}r_{2}^{*} & a_{\uparrow\downarrow}^{*}a_{\downarrow\downarrow}r_{1}^{*} & a_{\downarrow\uparrow}^{*}a_{\downarrow\downarrow}r_{2}^{*} & \vert a_{\downarrow\downarrow}\vert^{2} \\
\end{array}\right),
\label{eq:37}
\end{eqnarray}
where the decoherence factors $ r_{1}(t) $ and $ r_{2}(t) $ are given by
\begin{eqnarray}
r_{n}(t)=\prod_{k=1}^{N_{n}}(\vert \alpha_{nk}\vert^{2} e^{-i2\omega_{nk}t} + \vert\beta_{nk}\vert^{2} e^{i2\omega_{nk}t}).
\label{eq:38}
\end{eqnarray}
In general both expansion coefficients $ \alpha_{nk} $, $ \beta_{nk} $ and interaction strengths $ \omega_{nk} $
are random. For our purposes, we will assume that the baths are identical, which means we let expansion coefficients and
interaction strengths of the two baths be equal to each other as $ \alpha_{1k}=\alpha_{2k}=\alpha_{k} $,
$ \beta_{1k}=\beta_{2k}=\beta_{k} $ and $ \omega_{1k}=\omega_{2k}=\omega_{k} $. This assumption implies that the decoherence
factors of two baths are equal so that $ r_{1}(t)=r_{2}(t)=r(t) $. Thus, the reduced density matrix of two central spins is simplified to
\begin{eqnarray}
\rho_{c} =
\left(\begin{array}{cccc}
\vert a_{\uparrow\uparrow}\vert^{2} & a_{\uparrow\uparrow} a_{\uparrow\downarrow}^{*} r & a_{\uparrow\uparrow} a_{\downarrow\uparrow}^{*} r & a_{\uparrow\uparrow} a_{\downarrow\downarrow}^{*} r^{2} \\
a_{\uparrow\uparrow}^{*} a_{\uparrow\downarrow} r^{*} & \vert a_{\uparrow\downarrow} \vert^{2} & a_{\uparrow\downarrow}a_{\downarrow\uparrow}^{*} \vert r \vert^{2} & a_{\uparrow\downarrow}a_{\downarrow\downarrow}^{*} r \\
a_{\uparrow\uparrow}^{*}a_{\downarrow\uparrow}r^{*} & a_{\uparrow\downarrow}^{*}a_{\downarrow\uparrow}\vert r \vert^{2} & \vert a_{\downarrow\uparrow}\vert^{2} & a_{\downarrow\uparrow}a_{\downarrow\downarrow}^{*}r \\
a_{\uparrow\uparrow}^{*} a_{\downarrow\downarrow}(r^{*})^{2} & a_{\uparrow\downarrow}^{*}a_{\downarrow\downarrow}r^{*} & a_{\downarrow\uparrow}^{*}a_{\downarrow\downarrow}r^{*} & \vert a_{\downarrow\downarrow}\vert^{2} \\
\end{array}\right),
\label{eq:39}
\end{eqnarray}
where
\begin{eqnarray}
r(t)=\prod_{k=1}^{N}(\vert \alpha_{k}\vert^{2} e^{-i2\omega_{k}t} + \vert\beta_{k}\vert^{2} e^{i2\omega_{k}t}).
\label{eq:40}
\end{eqnarray}
We immediately observe that the form of $ \rho_{c} $ under the
assumption of identical baths is very similar to the form of the
output density matrix we obtained for classical noise Hamiltonian.
In particular, when the initial expansion coefficients  $ \alpha_{k}
$ and $ \beta_{k} $ are equal to each other, we will have exactly
the same form of the mapping obtained in Sec. \Rmnum{2}. Hence,
decay of $ r(t) $ to zero at later times and the form of the
possible Kraus operators in this case guarantee that the qubit
exchange symmetry properties of symmetric Bell states $ \vert B_{1}
\rangle $, $ \vert B_{2} \rangle $ and $ \vert B_{3} \rangle $
interacting with two local large spin environments will be the same
as their behavior under local stochastic noise fields. Since we interpret
decay of r(t) as a signature of decoherence, we identify decoherence as
the main source of spontaneous breaking of qubit exchange symmetry.

\section{Conclusion}

We examined the time evolution of exchange-symmetric Bell states for
local noise Hamiltonians having the same symmetry. For both
classical and quantum noise, we found a rather unexpected result
that not all Bell states preserved their symmetry. In fact, we
observed that exchange invariance property survived with a maximum
probability of $ 0.5 $ at the asymptotic limit. We conclude that
breaking of exchange symmetry for some possible final states is a
characteristic feature of decoherence.

\section{Acknowledgements}
This work has been partially supported by the Scientific and Technological Research Council of Turkey (TUBITAK) under grant 107T530.
The authors would like to thank C. Sa\c{c}l{\i}o\u{g}lu and T. Yildirim for helpful discussions, and comments.

\end{document}